\documentclass[%
 aip,
 amsmath,amssymb,
 reprint,%
]{revtex4-1}

\usepackage{graphicx}
\usepackage{dcolumn}
\usepackage{bm}

\usepackage[utf8]{inputenc}
\usepackage[T1]{fontenc}
\usepackage{mathptmx}

\usepackage{graphicx}
\usepackage{dcolumn}
\usepackage{bm}

\usepackage[utf8]{inputenc}
\usepackage[T1]{fontenc}
\usepackage{mathptmx}

\begin{document}

\preprint{AIP/123-QED}

\title[An apparatus for nondestructive and rapid comparison of mask approaches in defense against infected respiratory aerosols
]{An apparatus for nondestructive and rapid comparison of mask approaches in defense against infected respiratory aerosols}

\author{Donal Sheets}
 \email{donal.sheets@uconn.edu}

 \affiliation{%
Institute of Material Science, University of Connecticut, Storrs, CT USA, 06269
}%
\affiliation{ 
Department of Physics, University of Connecticut, Storrs, CT 06269
}

\author{Jamie Shaw}%
\affiliation{ 
Department of Physics, University of Connecticut, Storrs, CT 06269
}%

\author{Michael Baldwin}%
 \affiliation{%
Department of Diagnostic Imaging, University of Connecticut Health Center, Farmington, CT 06030
}%

\author{David Daggett}%
 \affiliation{%
Department of Molecular and Cell Biology, University of Connecticut, Storrs, CT USA, 06269
}%

\author{Ibrahim Elali}%
 \affiliation{%
Division of Nephrology, University of Connecticut Health Center, Farmington, CT 06030
}%

\author{Erin Curry}%
 \affiliation{%
Institute of Material Science, University of Connecticut, Storrs, CT USA, 06269
}%
\affiliation{ 
Department of Physics, University of Connecticut, Storrs, CT 06269
}

\author{Ilya Sochnikov}%
 \affiliation{%
Institute of Material Science, University of Connecticut, Storrs, CT USA, 06269
}%
\affiliation{ 
Department of Physics, University of Connecticut, Storrs, CT 06269
}

\author{Jason N. Hancock}%
 \email{jason.hancock@uconn.edu}

 \affiliation{%
Institute of Material Science, University of Connecticut, Storrs, CT USA, 06269
}%
\affiliation{ 
Department of Physics, University of Connecticut, Storrs, CT 06269
}

\date{\today}

\begin{abstract}
At the front lines of the world’s response to the COVID-19 pandemic are hero-clinicians facing a lack of critical supplies including protective medical grade breathing masks and filtering materials. At the same time, the general public is now being advised to wear masks to help stop the spread. As a result, in the absence of centrally coordinated production and distribution efforts, supply chains for masks, respirators, and materials for advanced filtration technology are immensely burdened. Here we describe experimental efforts to nondestructively quantify three vital characteristics of mask approaches: breathability, material filtration effectiveness, and sensitivity to fit. We focus on protection against water aerosols $>$0.3$\mu$m using off-the-shelf particulate, flow, and pressure sensors, permitting rapid comparative evaluation of these three properties. We present and discuss both the pressure drop and the particle transmission as a function of flow to permit comparison of relative protection for a set of proposed filter and mask designs. The design considerations of the testing apparatus can be reproduced by university laboratories and medical facilities and used for rapid local quality control of respirator masks which are of uncertified origin, monitoring the long-term effects of various disinfection schemes, and evaluating improvised products not designed or marketed for filtration.

\end{abstract}

\maketitle

\section{\label{sec:Introduction}Introduction:\protect\\}

SARS-CoV-2 is an infectious virus which is believed to be transmitted via respiratory aerosols\cite{Liu2020,WHO2020,NAP25769}. The threat of aerosol spreading is of such concern that the HVAC systems of hospitals are being re-engineered from their design intent so that infected air is released from the building immediately after potential infection in COVID wards\cite{CDC2019}. Near the peak of an infection curve \cite{Wehner2020}, the rate of infected patients can stretch the resources of hospitals and federal regulations have been adjusted to meet the realities of scarcity in the absence of enhanced production of personal protective equipment (PPE)\cite{Whalen2020}. This situation has set the medical profession on edge, and hospital systems in the northeast have so far responded with an all-hands-on-deck approach to the myriad challenges posed by COVID-19\cite{Anderegg2020,Schilling2020,Konda2020}. Independent researchers at local universities have also heard the call and moved rapidly to help their regional hospitals with the local problem of insufficient PPE such as face masks, face shields, and ventilators. Of these three vital hospital resources, face masks capable of blocking small virus-containing droplets require the most developed textile technology\cite{Tsai2020}. Disrupted supply chains have limited access to advanced textiles such as corona-charged melt-blown polypropylene, a material capable of supporting electrostatic filtration\cite{Kates2020}, which provides a high-efficiency, breathable, inexpensive and accessible mask certified as N95. With supply chains slowly recovering, the quality and safety of masks and respirators from new suppliers is critical to address.

Dwindling supplies of PPE in clinics and hospitals\cite{Khazan2020,KAPUST2020} combined with the concomitant emergence of appeals for the general population to begin donning masks\cite{Lamont2020} has strained the supply chain and led to local efforts to produce new technologies, as well as many proposed do-it-yourself mask approaches, in an effort to meet these needs \cite{Konda2020,CDCClothMasks2020,Bort2020}. In this environment, many claims of efficacy are made, while the data supporting such claims are often incomplete or lacking. Even in the case of certified respirators and surgical masks, historical variation in the testing regimes for particle filtering and infection control has been a complicating issue\cite{Rengasmy2017}. In addition, issues such as the presumed higher efficacy of N95 respirators over surgical masks in ultimately preventing infection remain controversial, with critical issues of fit and compliance likely confounding the utility of materials with ostensibly superior filtering properties\cite{Jefferson2020,Smith2016,Loeb2009,Leung2020}. In this time of crisis, accessible methodologies are needed which can rapidly compare novel materials and mask designs alongside currently certified materials. This document describes an experimental apparatus, devised and developed at the University of Connecticut and in collaboration with clinicians from the University of Connecticut Health Center during a short response period in March and April 2020. This apparatus is capable of making comparisons of essentially any proposed mask design with respect to filtration of water aerosol particles in the <1.0$\mu$m size range.

\begin{figure}
\includegraphics[width=.48\textwidth]{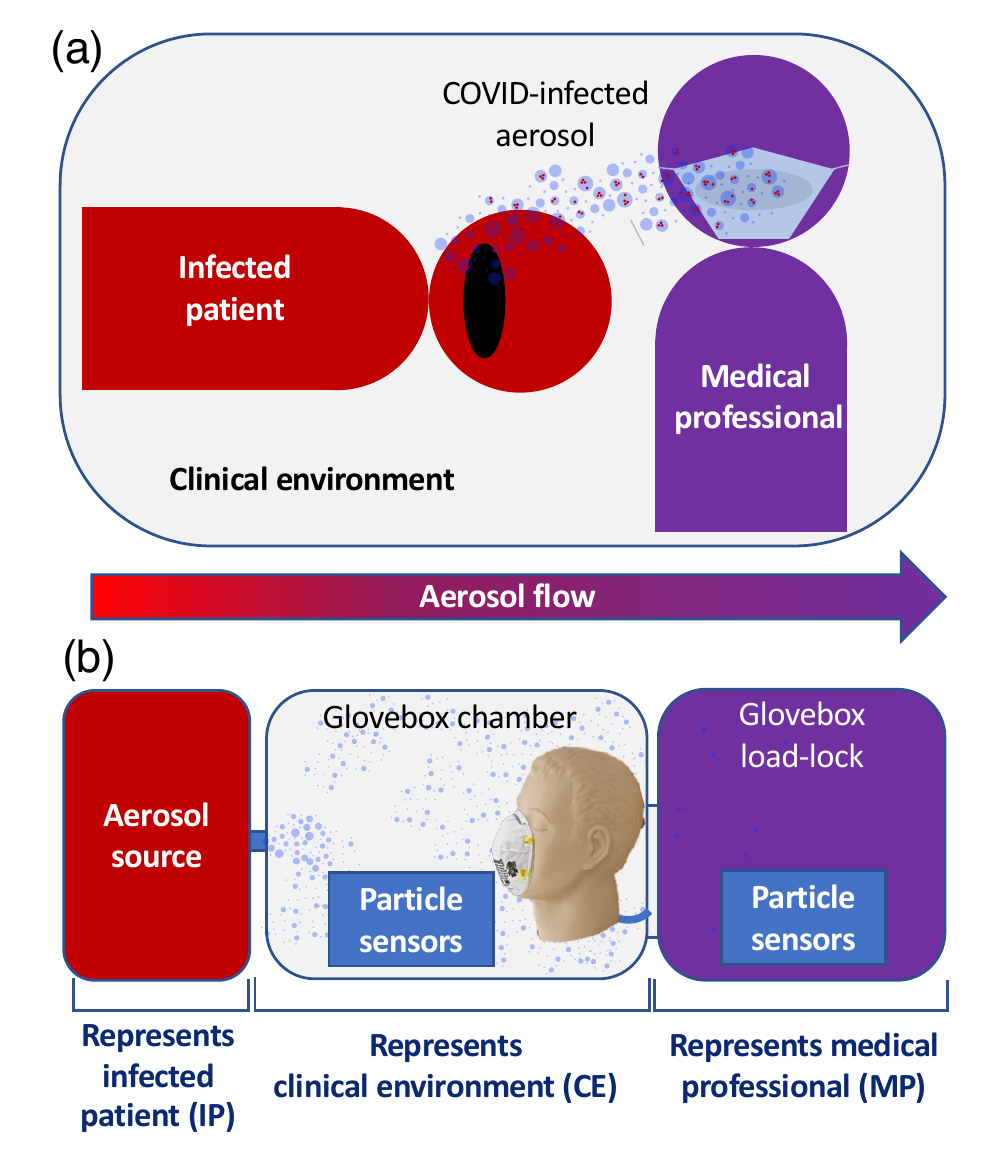}
\caption{\label{fig:1} Virtual clinical respirator mask test. (a) Schematic representation of the problem to be addressed. An infected patient (IP) exhales water infected aerosol into the clinical environment (CE). A mask is a filtration barrier protecting a medical professional (MP). (b) Schematic representation of the experimental apparatus. An aerosol source chamber represents the IP, a glovebox main chamber represents the CE, and the glove-box load lock represents the MP. Controlled aerosolized air flows through the system from left to right and pressures and particulates are measured for comparison studies at fixed environmental conditions. }
\end{figure}

\section{\label{sec:Experimental Apparatus}Experimental apparatus:\protect\\}
\subsection{\label{sec:Overview}Overview:}

\begin{figure*}
\includegraphics[width=150mm]{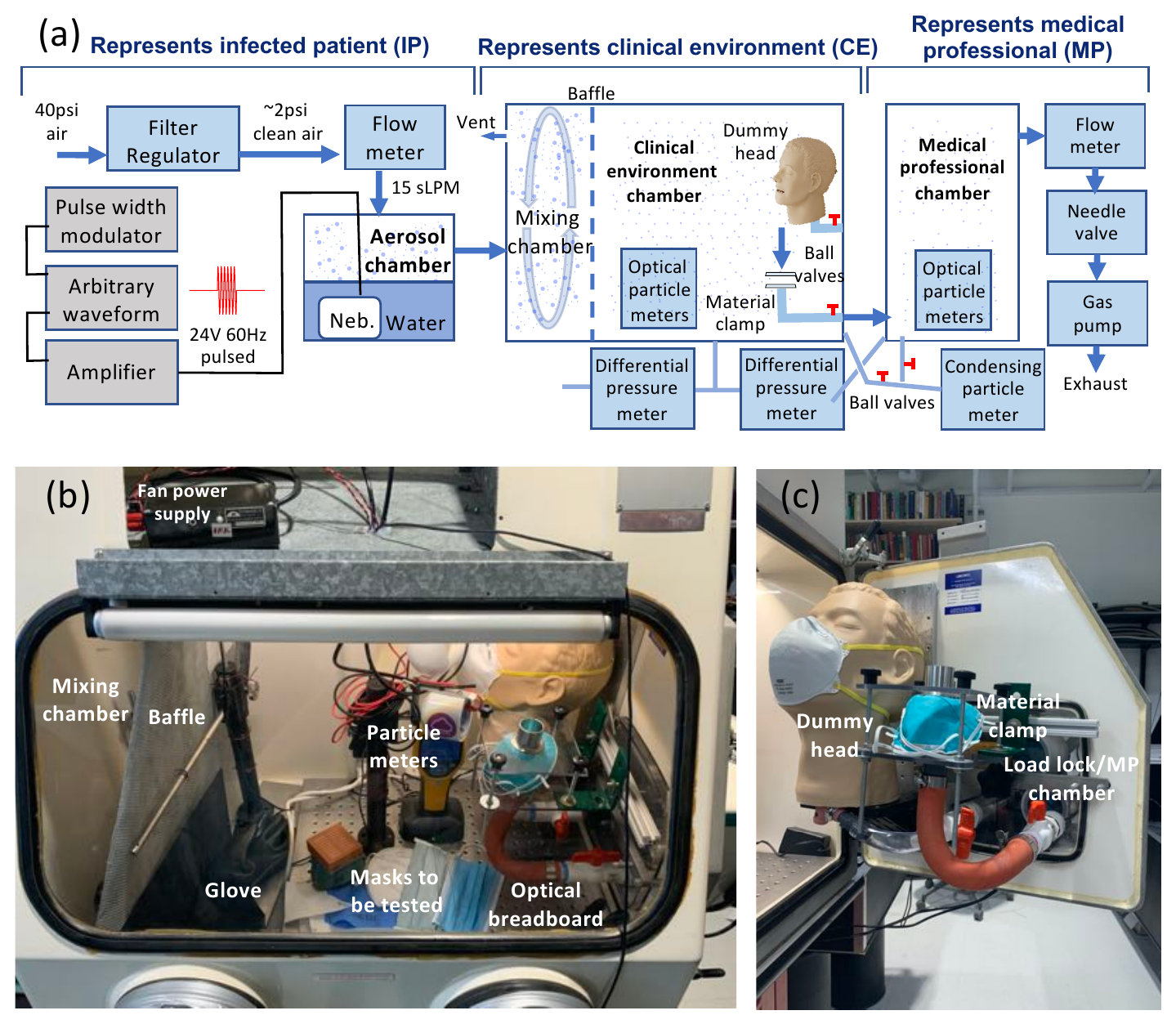}
\caption{\label{fig:2} (a) Schematic experimental setup with material tester engaged. Clean air is filtered and aerosolized by flowing over the surface of water containing a piezoelectric nebulizer (Neb.) element powered by 24VAC sine pulses arriving at 0.2Hz and variable duty cycle. The aerosolized air flows into a mixing chamber and through baffles, where it is drawn through a mask-donning dummy head (disabled in figure) or a material clamp tester and into the MP chamber. Optical and condensing particle meters measure humidity, pressure, temperature, and count particulate matter 0.02 to 10$\mu$m in diameter. (b,c) Photographs of the apparatus testing two N95 reference masks.}
\end{figure*}

Our approach relies on a simulation of the clinical situation\cite{Noti2012}, illustrated in Figure \ref{fig:1}a. Aerosol containing virus particles is exhaled by an infected patient (IP) into a clinical environment (CE), where it is potentially inhaled by a medical professional (MP). Each of these abbreviated terms corresponds to a separate chamber in our apparatus (Figure \ref{fig:1}b). Controlled flow of aerosolized air propagates from the IP chamber to the CE chamber and enters the MP chamber through either (a) a mask-donning dummy head or (b) a clamped-material tester depending on the state of two ball valves (Figure \ref{fig:2}c). The chambers are connected sequentially IP-CE-MP with gas flow fittings and feature calibrated sensors capable of measuring flow rate, pressure drops, and aerosol particle distribution in different size ranges. The two modalities of the apparatus permit assessment of (a) a mask on a dummy head with realistic respiratory geometry or (b) material-only properties regarding breathability and filtration efficiency over selected sections of a mask. Each of these modalities provide vital information to assess the practicality and effectiveness of protection against infectious diseases like those born by SARS-CoV-2.
The apparatus (Figure \ref{fig:2}) consists of a fiberglass glove box (Labconco 50350, Kansas City, Missouri, USA) with a sealable load-lock chamber mounted on a large swinging door which can be opened/closed and sealed quickly using toggle clamps. The large chamber of the glove box represents the CE where clinicians are exposed to potentially infected aerosol droplets exhaled by the infected patient (IP). A steady flow of aerosolized air flows from the IP through the CE and enters the pumped MP load-lock chamber through either (a) a fit tester dummy head (Laerdal Airway Management Trainer, Stavanger, Norway) or (b) a clamp-style materials tester based on Kwik-Flange (KF) vacuum fittings. Identical sets of particle meters in the CE and MP chambers monitor particulate sizes, humidity, temperature, and pressure on each side of the mask or material. Operation of two $\frac{3}{4}$in PVC ball valves enables either the (a) mask or (b) material tester, permitting various tests with the same chambers and sets of particle detectors. 
A gas pump, needle valve, and flow meter control and measure the flow through the mask or material and into the MP. Steady-state air flows into the CE from a controlled aerosolized air source, through the CE and enters the load-lock chamber representing the MP. Aerosol particle distributions are measured before and after the mask or the filter material and the experiment is repeated with an open system and a control high-performance mask (3M-1860 N95 standard) for comparison. Pressure drop is measured simultaneously using a differential pressure meter to assess breathability of masks and materials. Below we detail different components of the system following the air flow, from IP to CE to MP chambers.

\subsection{\label{sec:Aerosol Source}Aerosol Source:}
SARS-COVID-2 is a virus around 60-140nm in diameter\cite{Cascella2020} and is believed to be long-lived in airborne aerosol form, particularly in droplet nuclei, the dried-out residuals of droplets which potentially contain infectious pathogens \cite{Wells1955,Xie2007,Atkinson2020}. The half-life of SARS-CoV-2 in water aerosol is determined to be approximately 1 hour \cite{vanDoremalen2020}. While respiratory fluid is known to contain water as a dominant component, significant other constituents besides virions are common. A respiratory droplet 0.1nL in volume has a radius 2.8$\mu$m and a mass of 100ng and can contain as much as 1ng salt and 1ng protein. We simulate COVID-infected respiratory aerosols with a chamber where an ultrasonic nebulizer element submersed in water produces Faraday surface waves and cavitation\cite{Kooij2019}, emitting a plume of fine water aerosol. Much of the water evaporates and the fine particulate matter is determined by the residual ions. The use of water as a particulate has advantages for testing mask and material filtration specific to COVID-19. Water does not contaminate the system, compared to choices like saline solution aerosols, allowing for rapid screening of alternative PPE candidates, with direct comparison to desired outcome PPE, such as N95 respirators, within hours. 

The aerosol-generating subsystem used in the experiments presented here is outlined in Figure \ref{fig:2}a, left side. The nebulizer consists of an immersed piezoelectric element which is active when driven with a 60Hz 24V amplitude sine-wave. In order to achieve fine control of aerosol content of the air flowing through the system independently of the flow dynamics such as pressure drop and air currents, we pulse the nebulizer around 0.2Hz and variable duty cycle. This subsystem produces a plume of fine mist into the flow stream of the air once every five seconds, and the duration of the plume is controlled by varying the duty cycle of the input square wave. We find that after the mixing stage, there is no observable time structure in the aerosol content and the distribution is unaffected by the duty cycle $>$1\%. The measured relative humidity level in our main chamber is determined by ambient conditions and rarely exceeds 40\% relative humidity. In this way, fine control of aerosol content of the air flowing through the system is realized independently of the flow dynamics such as pressure drop and air currents. 

To achieve controlled 24V 60Hz pulse duration, we use an arbitrary waveform generator (AWG; Instek 2000 series, Taiwan) which is amplified by a waveform amplifier (Accel TS250, USA) to increase the current supplied to drive the element. If the nebulizer is on, we find that 45 seconds of operation completely saturates the detectors in the CE ($>$6.5$\times 10^4$ particles/sec $>$0.3$\mu$m). We gate the AWG with a square wave gate pulse with variable duty cycle from a function generator (Instek 2000 series, Taiwan). Alternatively, variable duty cycle of AC power with low-cost electronics could be achieved using a short period repeat cycle timer or a solid-state relay operated via a low-cost function generator. 

Aerosol was sampled in the CE chamber using an optical particle detector (Fluke 985 clean room particle detector, USA) and found to have a particle distribution mostly below 1$\mu$m, consistent with prior reports on a similar device\cite{Kooij2019} after significant evaporation. Aerosol in this range is both challenging to filter using woven or common fabric materials but also makes up the vast majority of exhaled particulates\cite{Smolik2001} so is suitable for our apparatus.

\begin{figure*}
\includegraphics[width=150mm]{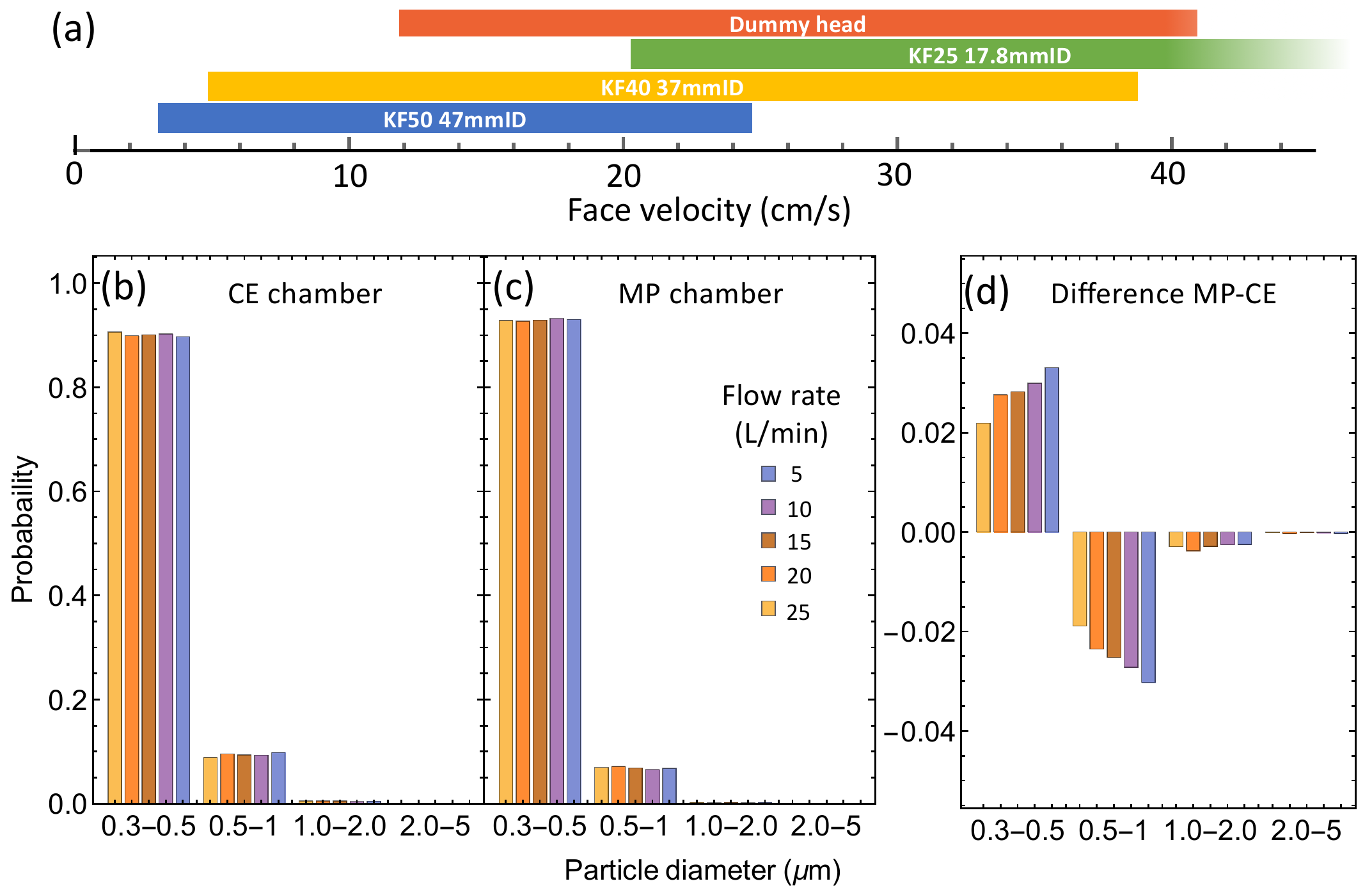}
\caption{\label{fig:3} (a) Available face velocities using the various fitting attachments. The flow range used in this study was 5-25L/min but the variable area permits a wide range of face velocity spanning the most common certification standards in the range of 9 to 25cm/s\cite{Rengasmy2017}. Particle distributions in the (b) CE and (c) MP chambers, normalized by the total number of particles in each. (d) shows the difference MP-CE and highlights a shift in the distribution to smaller particulates, consistent with a small degree of evaporation.}
\end{figure*}

\subsection{\label{sec:Mixing chamber}Mixing chamber:}
Once the aerosolized air is pushed into the CE glovebox chamber, two DC motor fans, arranged to set up a vertical circulating flow pattern on the left end of the glovebox chamber, rapidly mix the aerosol and prevent settling. During this mixing process, large droplets evaporate and approach an equilibrium size distribution. Applying the standard theory for settling time and terminal speed under Stokes-law drag force developed for spherical aerosol droplets in air, we estimate the settling velocities of the 0.3 to 1$\mu$m water aerosol observed in our chamber to be in the range 1 to 10 cm/h, which implies that the aerosol in the chamber would stay well mixed over a few hours, more than sufficient to reach the mask and materials testers in the CE chamber. Solutes such as NaCl or KCl can be added to the water in the IP chamber and control the particulate size distribution as well as concentration.

The mixed aerosol crossed through a baffle made from two layers of steel mesh with a 3.5mm grid pattern. Using a hot wire anemometer, the measured air speed in the mixing chamber is of order 100cm/sec but less than 1cm/sec on the downstream side of the baffles. For a circular flow pattern of the mixing chamber has approximately 1m circumference, we estimate the chamber churns the air 5 times per aerosol pulse delivered 5 times/sec, consistent with the lack of detectable time structure has been observed in the CE 
chamber.

\subsection{\label{sec:Particle detectors}Particle detectors:}
Aerosolized air was sampled on the right side of the CE chamber using optical particle detectors before transiting the mask or material into the MP chamber where a matching set of particle meters measured the aerosol distribution for comparison. Three different types of meters are employed to span different size distribution ranges.

Fast readout particle detectors developed by PurpleAir (Utah, USA) are based upon the PLANTOWER 5001 (Beijing Plantower, China) optical meters. In collaboration with PurpleAir, we have gratefully acquired test software permitting complete readout and logging of the two detectors per unit (two units/chamber) at approximately one second intervals. These meters have been studied under various conditions and correlate reasonably well with higher-performing and more costly particle meters \cite{LevyZamora2019}. They do not work as well in high humidity environments ($>$50\%), are known to have lowered efficiency (~\%50) in the finest particle channel (0.3-1.0$\mu$m) and are designed to perform well at high particle concentration levels. These meters are used during the experiment to make decisions during data acquisition as well as in analysis.

NIST-traceable cleanroom Fluke 985 integrating and logging optical particle detectors are placed in each chamber during a typical run. These detectors are best suited for lower particle concentrations. The devices are programmed to collect 15 seconds with 2 minute wait times between.

A single P-Trak 8525 ultrafine condensing particle counter is used to detect integrated particle count in the 0.02-1.0$\mu$m range. This costly device consumes reagent grade isopropanol and is placed outside the chamber and connected to a small ball valve manifold with equal-length tubing to the CE and MP chambers. A single condensing particle counter was also used and a small ball valve manifold permitted fast switching between CE and MP chambers.

\subsection{\label{sec:Mask Tester}Mask tester:}
A dummy head was removed from a Laerdal Airway Management Trainer, designed to give an accurate representation of human airflow pathways. The lungs were removed and the right main bronchus was closed with a rubber stopper and hose clamp. The left main bronchus was also hose-clamped onto a $\frac{1}{2}$ in PVC barb fitting to to $\frac{3}{4}$ in NPT pipe thread, a short segment of vinyl tubing, and another barb into a PVC ball valve which is open when the mask tester is in use. The valve feeds through an acrylic door to the load-lock chamber using PVC pipe fittings and seals with an o-ring. Optical breadboard sections mounted inside the door and in the bottom of the CE chamber provide secure mounting options for the particle sensors as well as dummy head and materials tester.

\subsection{\label{sec:Materials tester}Materials tester:}
\begin{figure*}
\includegraphics[width=130mm]{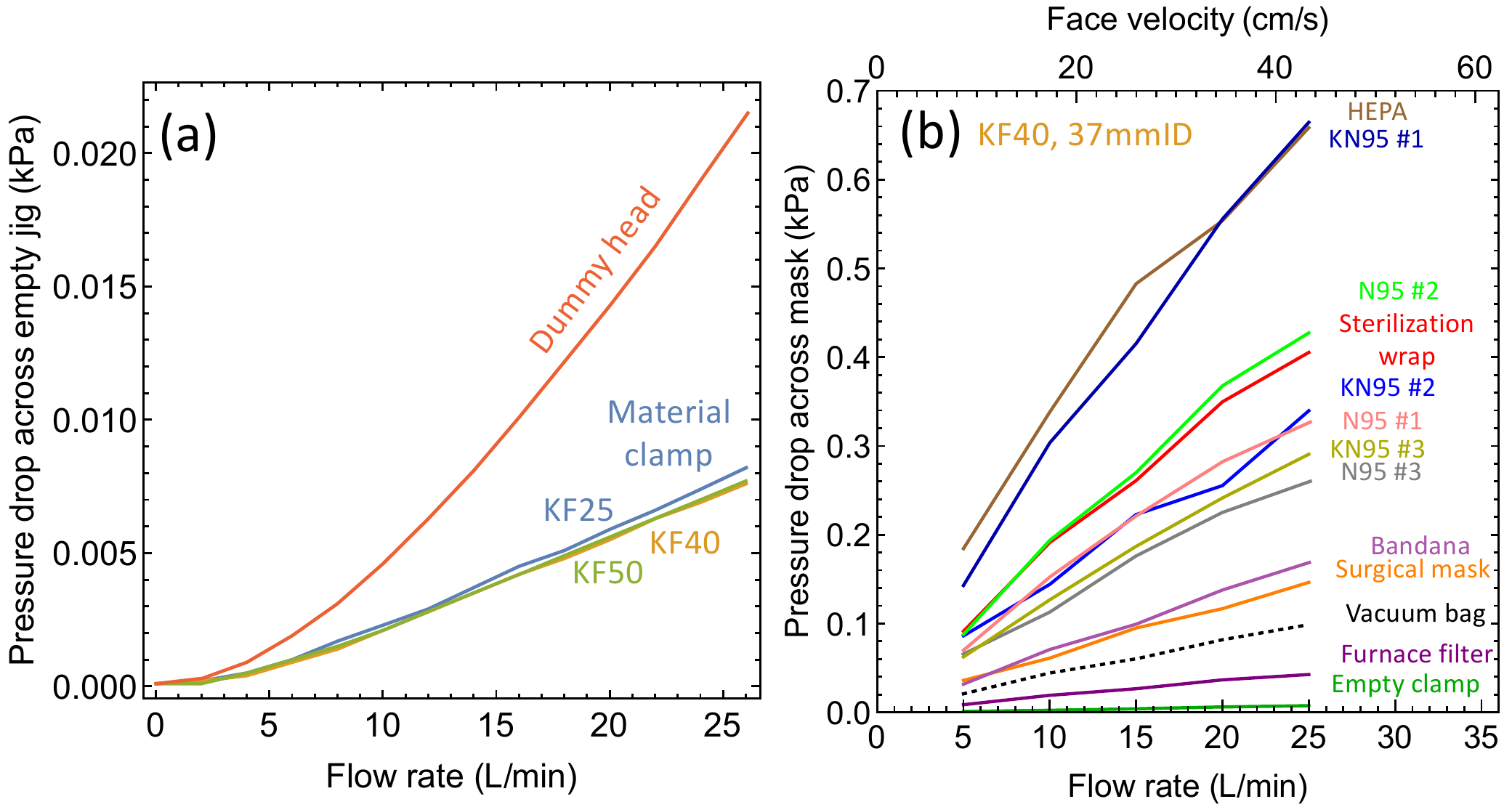}
\caption{\label{fig:4} Breathability measurements. (a) Pressure drop across a dummy head wearing no mask and a materials clamp with various vacuum fittings sizes, demonstrating the sensitivity of the pressure drop measurement. (b) Pressure drop versus flow across some filters, masks, and filter material candidates in the materials tester using the KF40 fitting. A pressure drop spread between quality and mediocre masks is substantial, making some masks dangerous to use if high flow impedance draws contaminants through leaks and gaps.}
\end{figure*}

To rapidly identify whether the fit or material are the weak link in any proposed mask design, we have enabled an option to disable the dummy head and open the CE-MP chamber connections though a dedicated materials tester. This jig is made active by opening a PVC ball valve connecting the MP chamber through PVC hard line or thick-wall $\frac{3}{4}$” rubber tubing terminating in a KF25 vacuum fitting pointing upwards and held fixed in the center of a 6”x6” polycarbonate plate mounted to the glovebox door. Four $\frac{1}{4}$”-20 threaded rods in each corner of the polycarbonate plate form a materials clamp with a matching plate above and a second KF25 fitting. In order to access lower face velocities, we can introduce one of two conical adapters (KF25 to KF40 or KF25 to KF50) and matching vacuum fittings (KF40 or KF50, respectively) to distribute the flow over a larger area. Figure \ref{fig:3}a summarizes the parameters achievable with each of these fittings and Figure \ref{fig:4}a shows the transmission and pressure drops over these choices of fitting. This configurability permits access to face velocities up to 170cm/sec at a flow 25L/min with a 17.8mm hole on the KF25 fitting and down to the lower value of 2.9cm/sec at 5L/min flow with the 47mm hole in the KF50 fitting. Our system could be operated at higher flows with suitable replacement of mass flow meter with a high flow mass flow meter or controller.

Both jigs were mounted inside the glovebox door as shown in Figure \ref{fig:2}c in order to permit short tubes in geometries which do not move when the door is opened. To facilitate this mounting, a small strip of optical breadboard was bolted to the inside of the glovebox door to present suitable and versatile mounting options. The material tester can be accessed using the box gloves enabling rapid swapping of the material without the need to wait for the chambers to equilibrate. The ability to swap out masks consecutively without altering environmental conditions is helpful for direct comparison of material approaches.

\subsection{\label{sec:Particle Distribution}Particulate distribution:}
Experiments at various conditions indicate that pure water aerosol <1$\mu$m in diameter evaporates quickly, reaching its equilibrium size in less than one second\cite{Cascella2020,Smolik2001}. Figure \ref{fig:3}b,c show the normalized size distribution of aerosol measured in the MP and CE chambers, showing that for tap water aerosol source, $>$90\% of the observed particles are in the finest size bin 0.3-0.5$\mu$m and there is only a weak redistribution of sizes on transit from the CE to MP. Figure \ref{fig:3}d shows the difference of these plots, (MP minus CE) and reveals a small but detectable redistribution of the aerosol before and after the mask or material testers. The shape of the distribution changes no more than 3.3\% between bins and does not significantly alter our comparative filtration measurements of masks and materials. Note that the difference depends only slightly upon flow, with greater change occurring for lower flow as expected from evaporation. We estimate for the slowest flow rates used here (5L/min), aerosol is drawn from one to another in less than 3 seconds.

In addition to changes in the shape of the distribution, there is also a flow-dependent transmission of the total number of aerosol particles $>$0.3$\mu$m through an open pipe. This arises from a well-known effect of aerosol collisions in the connecting pipe\cite{Kumar2008} - the longer the aerosol spends in the pipe, the more chance for a collision with a pipe wall and removal from the flow stream. As expected, we observe the lowest transmission of an empty pipe for the lower flows. To isolate the transmission of the mask or material, it is therefore necessary to address the loss of particles in the system when there is no mask present. For each measurement, the number of particles $N$ are measured for the MP and CE chambers with a mask in the clamp ($N_{m,MP}$,$N_{m,CE}$) as well as the same for an open pipe or dummy head ($N_{o,MP}$,$N_{o,CE}$). To account for this loss, we divide the observed transmission $N_{m,MP}/N_{m,CE}$ of a given mask/material by a reference measurement of the open jig $N_{o,MP}/N_{o,CE}$ at a given flow, permitting isolation of the transmission T of the mask or material: $T$ = $100 \times N_{m,MP} N_{o,CE} /N_{m,CE}N_{o,CE}$ and the filtration efficiency is $f$ = $100-T$. We emphasize that our goal is a comparison of mask approaches to one another rather than an absolute certification standard.

\subsection{\label{sec:Flow Control}Flow Control:}
A gas pump (Oxford GF10, UK), needle valve, and thermal mass flow meter (TSI 4100, USA) connected to the MP chamber sustains constant and controllable flow of gas through the system. The gas pump is located in a separate pump room, connected by $\frac{1}{4}$in tubing to a constrictive needle valve which permits fine control of the flow through the system. A thermal mass flow meter and a protective inlet filter recommended by the manufacturer is installed upstream and sensitively reads the mass flow with precision of 0.001L/min. Feedback mass flow control could be introduced easily although we find the drift in flow is quite small.

\begin{figure*}
\includegraphics[width=130mm]{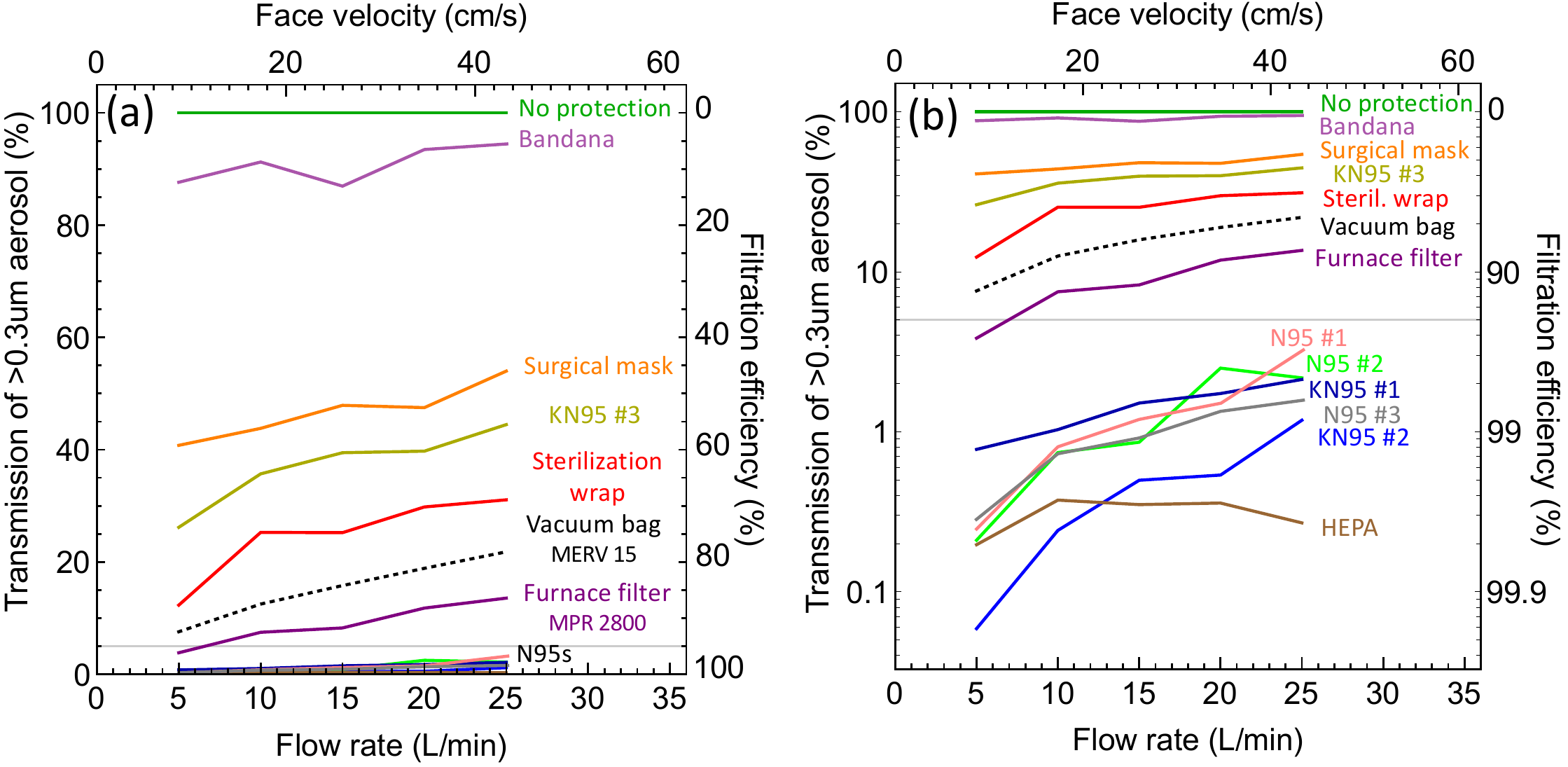}
\caption{\label{fig:5} Relative transmission measurements. (a) Linear (b) log plot of normalized transmission of select filter materials through the 37mm ID KF40 flange attachment. Most of the KN95 or N95 mask material we tested comply with the transmissions specifications. Alternative filtration media were not close enough to the 5\% threshold (gray line) to suggest its widespread use as a replacement to high-grade material.}
\end{figure*}

\section{Experimental results}

\subsection{\label{sec:Breathing}Breathing, pressure drop, flow, face velocity:}

Crucial for high performance masks is that low effort is required to inhale or exhale through a mask or mask material. This effort is typically quantified using the pressure drop at a given flow rate. The flow requirements on breathing are given by the situation, which for humans at rest is around 5-6 L/min on average assuming a tidal volume of 0.5L cycled 10-12 times per minute. Time-resolved measurements of human respiration give flow rates in the range 10 L/min peak exhalation, with larger inhalation flow rates of order 25L/min. Exertional breathing in healthy humans can reach peak inhalation flow exceeding 600L/min. The National Institute for Occupational Safety and Health (NIOSH) performs tests of mask and mask materials for N95 certification at 85L/min, where $>$95\% of a defined distribution of NaCl particles $>$0.3$\mu$m are blocked\cite{Tsai2020}.

For any mask the flow Q is distributed over the active filter material area A with a mask-averaged speed called the face velocity v = Q/A. Some standards for mask testing like NIOSH, specify flow, while others such as FDA specify face velocity\cite{Rengasmy2017}. For the NIOSH test over a typical half face respirator such as the 3M-1860 of area 150cm$^2$, the mask-averaged face velocity is around 9.3cm/sec while the same exercise for a surgical mask of area $^2$ results in a face velocity of 14.2cm/sec\cite{Rengasmy2017}. Processes behind filtration include inertial impaction and diffusion in addition to electrostatic effects in the case of N95\cite{Tsai2020,Rengasmy2017} and this speed of impact determines the filtration efficiency. In what follows we will express the flow rate as well as face velocity, the latter of which permits comparison of filter measurements which used different areas. We emphasize that use of variable size vacuum fittings permits wide changing of the face velocity for a given flow range as shown in Figure \ref{fig:3}a.

\subsection{\label{sec:breath}Pressure drop over typical filter materials:}
Breathing through a mask is always more effort than breathing without one. The resistance to flow of the mask is determined by the filter material and the fit of the mask and can be quantified by the pressure drop at a given flow. In our experiments, the pressure difference between the MP and CE chambers is the pressure drop over either the mask or material being tested. We measure the pressure drop using a differential pressure meter (TSI 9565, Minnesota, USA) connected to each chamber by tubing of equal length.

Figure \ref{fig:4}a shows the pressure drop versus flow for the dummy head and each of the KF fittings attached to the materials clamp. These constrictions produce a measurable pressure drop much smaller than that of typical masks and materials. Figure \ref{fig:4}b shows the pressure drop over a set of materials with the KF40 vacuum fitting in the materials clamp. We note in particular the spread of pressure drops among masks labelled N95 and KN95. Large pressure drops across filter material can significantly reduce the protection of a given mask, since the required air flow is distributed over both the mask and leaks. At a given flow, as the pressure drop across the mask increases, the draw of contaminants through leaks and facial fit gaps increases. We return to this point below in section \ref{sec:fit}.

\subsection{\label{sec:filter}Filtration efficiency of filter material:}
Figure \ref{fig:5}a shows the transmission of all particles $>$0.3$\mu$m versus flow rate for certified N95s and many other materials using the KF40 materials tester attachment. The woven bandana provides very little protection from aerosol in the size range of our experiments, around 90\% of particulates counted were in the 0.3-0.5$\mu$m range. It does however have an appreciable pressure drop, around 60\% of an N95. A surgical mask has far superior 50\% filtration efficiency compared to a bandana but is similarly breathable with similar pressure drop at the same flow rate. Sterilization wrap has been proposed as a highly efficient filter candidate for improvised PPE approaches\cite{THOMPSON2020} and sterile and readily available in many clinical settings. Compared to a surgical mask, the material has higher filtration efficiency , but it is less breathable. While a higher filtration aid protection, a higher pressure drop means that to sustain a given flow, the pressure across leakages and draw across soft points in the facial seal will support more flow which is detrimental to protection. Another class of PPE filter materials are HEPA-type vacuum bags and electrostatic furnace filters. These have standardized filtration rating systems of MERV and MPR and examples are shown to achieve very good filtration and excellent low pressure drop. Multiple layers of materials in this class could give excellent filtration in an abundant material source alternative to melt-blown polypropylene, the finest filtration component of the electrostatic N95\cite{Tsai2020}.

The logarithmic scale in Figure \ref{fig:5}b brings out the low-transmission region of interest for comparison of N95s for quality control and inspection of deterioration effects from disinfection schemes on post-cleaning filtration efficiency. All N95, most KN95s, and a high-efficiency based upon PFTE-coated PET show $>$95\% filtration efficiency. This last material excels at filtration but is far too difficult to breath through. Note that one uncertified mask labelled KN95 \#3 performs poorly with only ~60\% filtration efficiency. While this mask has filtration performance slightly better than a surgical mask, the pressure drop is greater which means more flow is expected through the small leaks around the mask and may not in fact be safer in practice. This observation highlights the importance of local testing of mask and materials of uncertified origin as well as the need to assess both filtration and pressure drop properties of materials. 

\subsection{\label{sec:fit}Mask facial fit:}
\begin{figure}
\includegraphics[width=70mm]{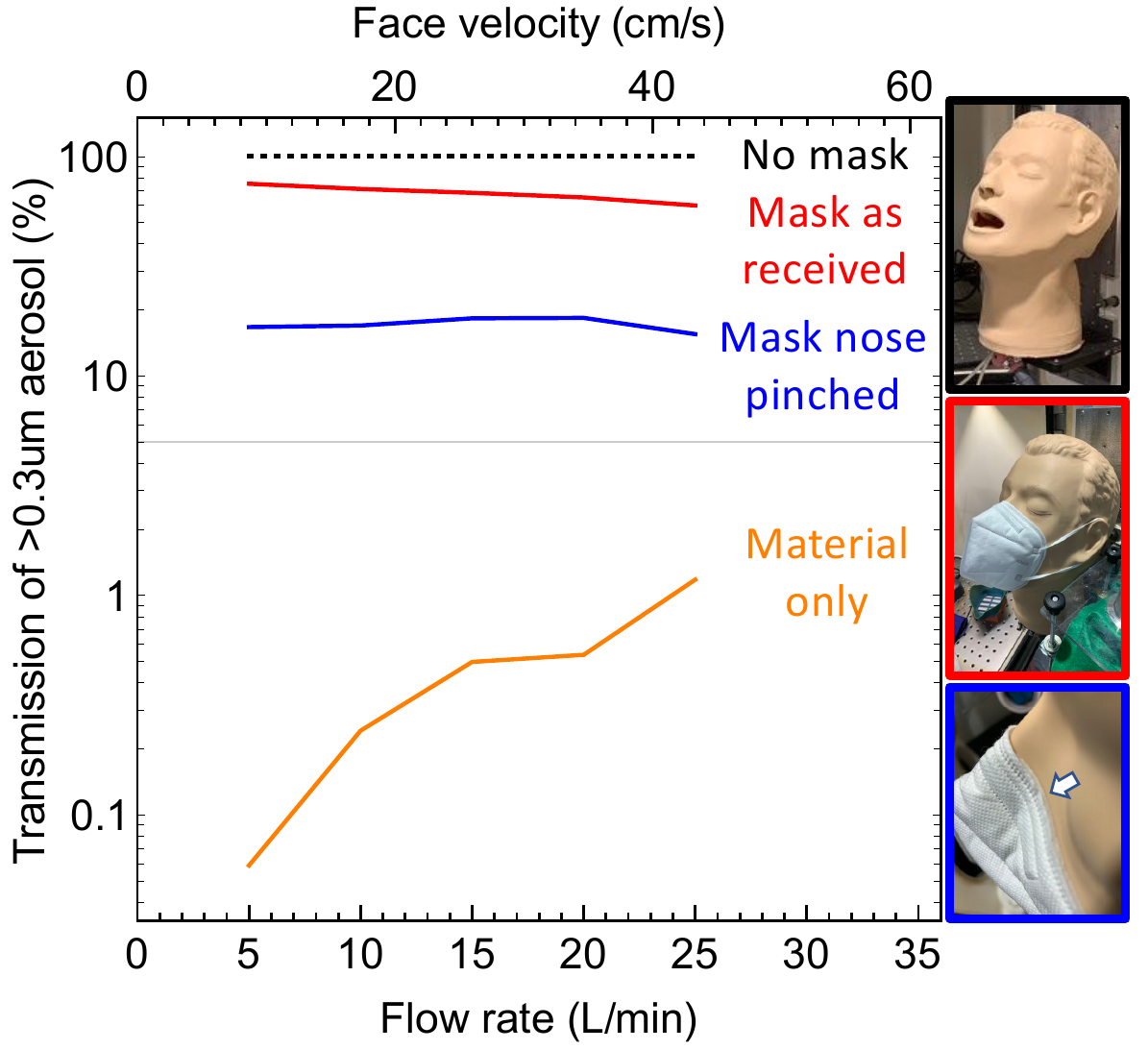}
\caption{\label{fig:6} Mask fit result. A KN95 mask is donned and transmission measured before and after the nose was pinched. These results emphasize the importance of facial fit in effective protection.}
\end{figure}

Sections \ref{sec:breath} and \ref{sec:filter} describe materials tests of pressure drop and filtration efficiency for materials with issues of facial fit completely isolated. Inclusion of a second test channel through the dummy head permits an assessment of mask fit and its potentially deleterious contribution to protection. To expose this sensitivity, Figure \ref{fig:6} shows the transmission of aerosol through the dummy head with a mask labelled KN95 fitted as received and with the metal nose piece pinched. The use of a glove box is a necessary feature for making small in-situ adjustments to study the facial fit sensitivity. The drastic improvement from $>$50\% transmission to $<$20\% transmission highlights the benefits of fitment. At the same time, the degree of protection is still much less than the material alone, which transmits less than 1\%. This demonstrates the apparatus ability to rapidly compare developing mask approaches but is limited in addressing dynamic factors such as head and jaw movement while talking as well as variation of head forms.

\section{\label{sec:Conclusions}Conclusions:\protect\\}
We have demonstrated the construction and evaluation of a fast-turnaround apparatus capable of evaluating three important characteristics of aerosol-filtering masks: breathability quantified through pressure drop, filtration quantified through particulate transmission, and facial fit as determined from direct measurements on an anthropic head form. We have demonstrated the sensitivity of the measurements and considered sources of error in the measurement. We present comparison of high-technology electrostatic filters, surgical masks, instrument wrap fabric, and common textiles against complete lack of protection. The glovebox-based design concept has many advantages such as rapid screening of many masks without disrupting aerosol generation and flow. The design can be implemented with resources common to many universities and hospitals and can be implemented with low-cost sensors and electronics to screen masks that may vary substantially from dangerously poor to very high performance depending on the quality of the materials used.
\begin{acknowledgments}
This work would not be possible without the support of Eleiza Braun, Michael Bailey, Wesley Byerly, Ray Celmer, Jason Chandler, Tamara Corioso, Terri Dominguez, Mike Jednak, Andrew Kelly, Dave Perry, Cindy Polinksi, and Barrett Wells. The authors would like to thank Drew Gentner, Kevin Hsueh, Barbara Mellone, Lisa Lattanza, Shawn London, Michael Plumley, Katherine Schilling, Kristina Wagstrom, Ben Waxman, Whitney Waxman, Larry Wilen and Christopher Wiles for valuable conversations. We are especially grateful to Adrian Dybwad and PurpleAir for support in designing the software acquisition system. We would like to also like to thank Joseph Luciani, Matthew Phelps, Amani Jayakody, Asanka Amarasinghe, Kaitlin Lyszak for assistance and for supporting the work. This work was supported in part by National Science Foundation Grant No. NSF-DMR-1905862.
\end{acknowledgments}

\section{Data Availability}

The data that support the findings of this study are available from the corresponding author upon reasonable request.

\begin{appendix}

\section{Materials tester detail}

\begin{figure*}[h]
\includegraphics[width=\textwidth]{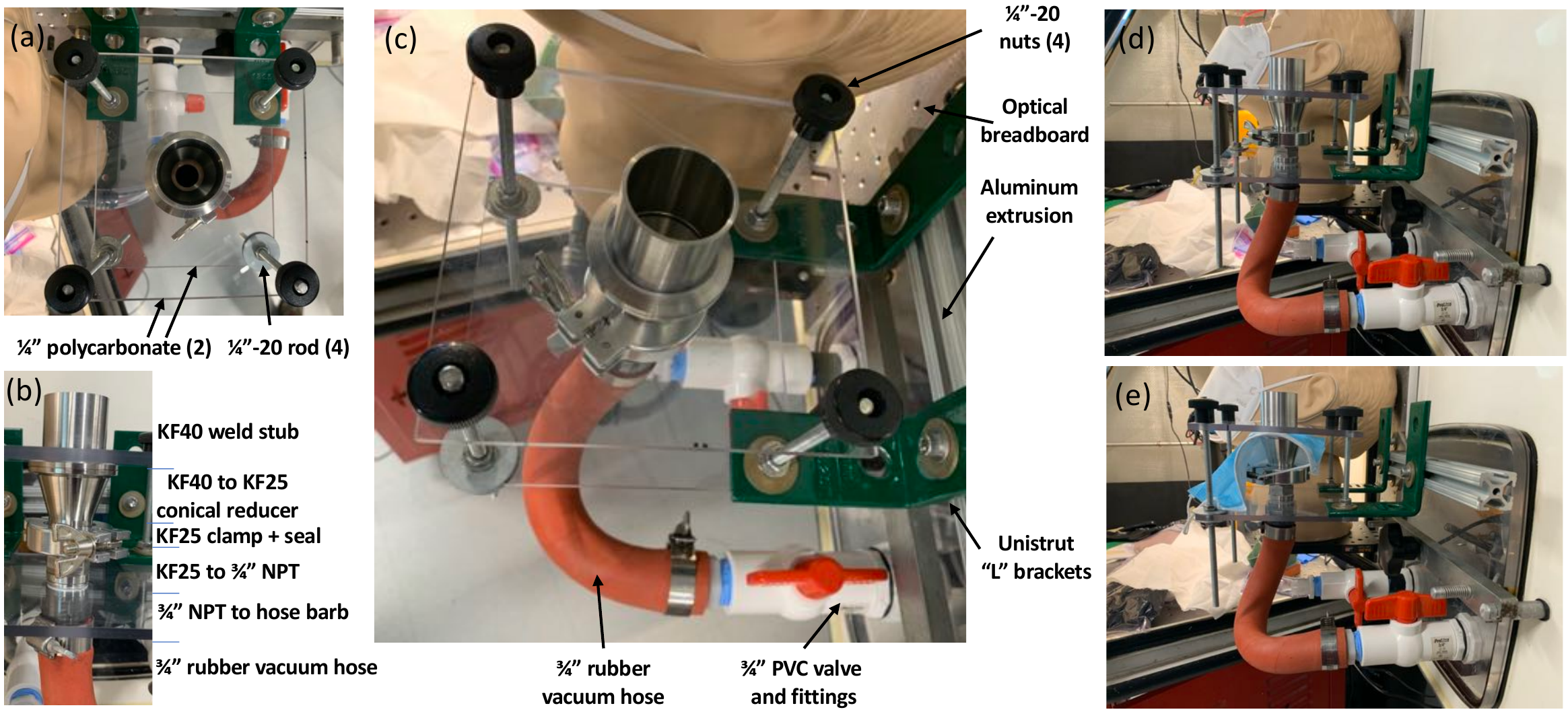}
\caption{\label{fig:7} Material tester detail. (a-c) Stainless steel vacuum fittings, polycarbonate sheets, and threaded rod are used to make a clamp for materials tests. (d) Empty material tester. (e) Nondestructive testing of a surgical mask.}
\end{figure*}

\begin{figure}[h]
\includegraphics[width=70mm]{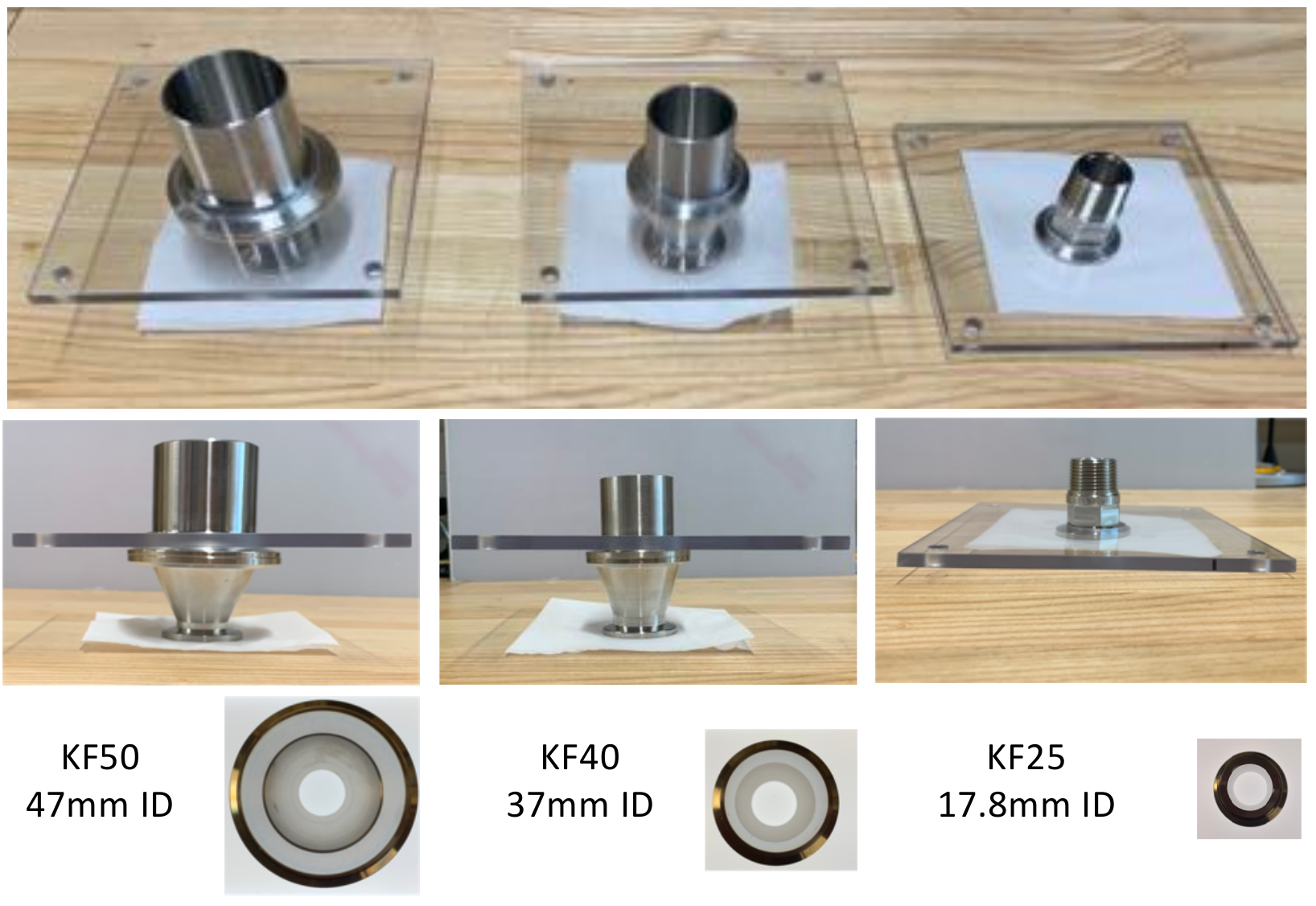}
\caption{\label{fig:8} Three sizes of KF flange permit adjustment of the face velocity for a given flow range.}
\end{figure}

Figure \ref{fig:7} details the materials tester mechanism. A flow-through clamp is based upon Kwik-Flange vacuum fittings to create a tight and reproduceable clamp on thin pieces of material without invasive handling or destruction of a mask or material being tested. The clamp can be operated with the gloves of the glovebox, permitting direct comparison of materials with each other or with an empty holder under the same environmental conditions. In order to ensure the mating flanges clamp parallel to one another and form a tight uniform seal on the material, the KF flanges are compressed between two polycarbonate plates using $\frac{1}{4}$-20 threaded rod as shown. Large knobs on the clamping nuts are convenient for changing masks while using the gloves. An advantage of the KF approach is the possibility to change the filter or mask area being probed which leads to a change in the tested face velocity range as illustrated in Figure \ref{fig:3}a of the main text. Details of the clamp mechanism options for the three sizes available for the materials tester are shown in Figure \ref{fig:8}.

\section{Direct face velocity measurements}

\begin{figure}[h]
\includegraphics[width=70mm]{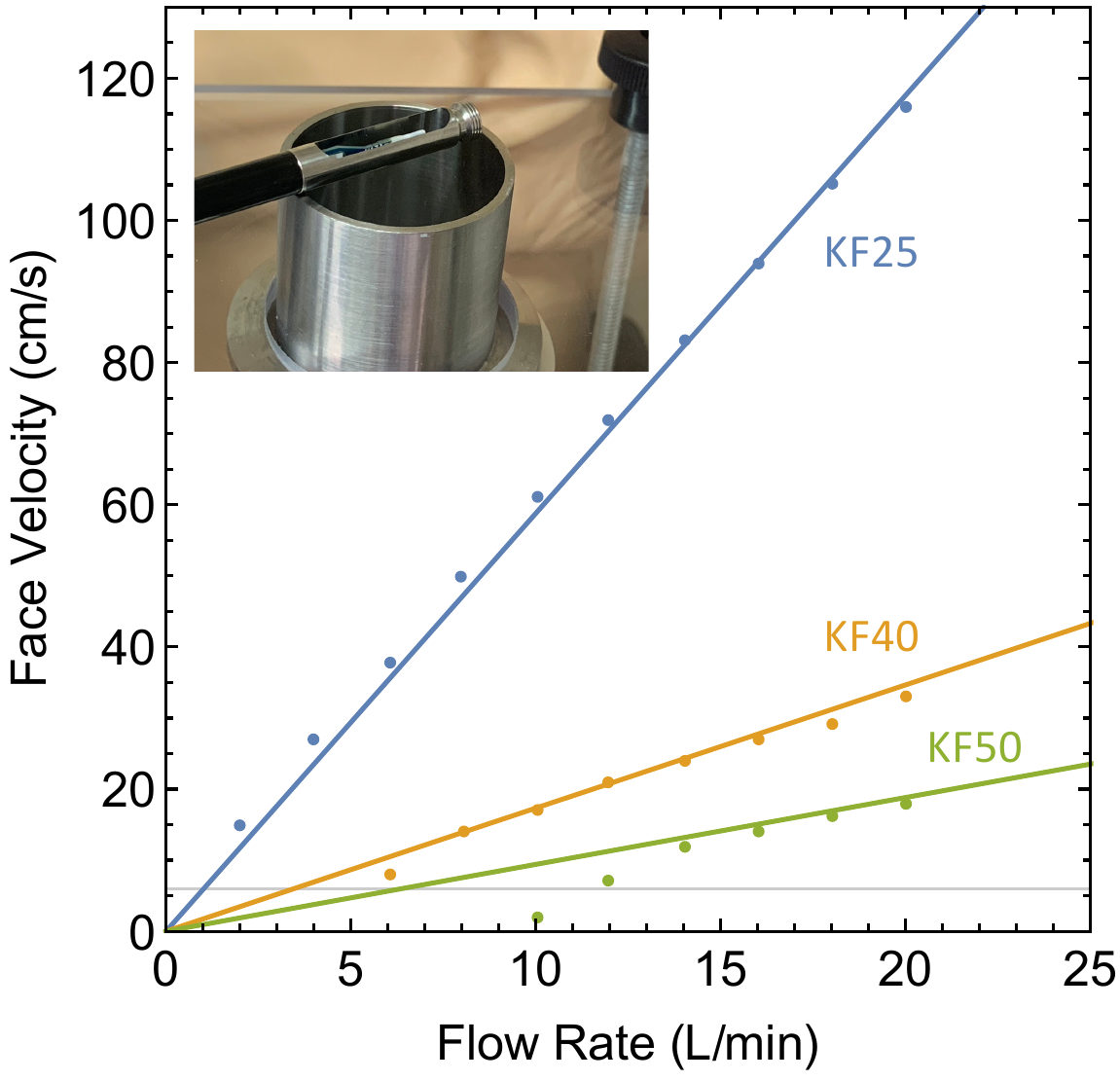}
\caption{\label{fig:9} Results of the anemometer face velocity measurement (points) along with the volumetric flow divided by KF weld stub inner area (lines).}
\end{figure}

Figure \ref{fig:9} shows results of an experiment using a hot wire anemometer to directly measure the face velocity across the open weld stub where air enters the materials tester. The lines through the points are the volumetric flow rate divided by measured area. This plot shows that our face velocities are as expected and that there are no significant leaks in the MP chamber.

\end{appendix}

\section{References}
\nocite{*}

\bibliography{masktest}

\end{document}